\documentclass[aps,prb,twocolumn,superscriptaddress, longbibliography]{revtex4-1}

\usepackage{graphicx}
\usepackage{amsmath,amsfonts,amssymb}
\usepackage[colorlinks=true,linkcolor=blue,citecolor=blue]{hyperref}

\usepackage{layouts}

\usepackage[separate-uncertainty = true,multi-part-units=single]{siunitx}
\usepackage{longtable, tabu}
\usepackage{booktabs}
\usepackage{multirow}
\usepackage{tabularx}

\begin{document}


\title{Out-of-plane corrugations in graphene based van der Waals heterostructures} 



\author{Simon Zihlmann}
\email{s.zihlmann@unibas.ch}
\thanks{These authors contributed equally.}
\affiliation{Department of Physics, University of Basel, Klingelbergstrasse 82, CH-4056 Basel, Switzerland}

\author{P\'eter Makk}
\email{peter.makk@mail.bme.hu}
\thanks{These authors contributed equally.}
\affiliation{Department of Physics, University of Basel, Klingelbergstrasse 82, CH-4056 Basel, Switzerland}
\affiliation{Department of Physics, Budapest University of Technology and Economics and Nanoelectronics 'Momentum'
Research Group of the Hungarian Academy of Sciences, Budafoki ut 8, 1111 Budapest, Hungary}

\author{Mirko K. Rehmann}
\affiliation{Department of Physics, University of Basel, Klingelbergstrasse 82, CH-4056 Basel, Switzerland}
\affiliation{Swiss Nanoscience Institute, University of Basel, Klingelbergstrasse 82, CH-4056 Basel, Switzerland}

\author{Lujun Wang}
\affiliation{Department of Physics, University of Basel, Klingelbergstrasse 82, CH-4056 Basel, Switzerland}
\affiliation{Swiss Nanoscience Institute, University of Basel, Klingelbergstrasse 82, CH-4056 Basel, Switzerland}

\author{M\'at\'e Kedves}
\affiliation{Department of Physics, Budapest University of Technology and Economics and Nanoelectronics 'Momentum' Research Group of the Hungarian Academy of Sciences, Budafoki ut 8, 1111 Budapest, Hungary}

\author{David Indolese}
\affiliation{Department of Physics, University of Basel, Klingelbergstrasse 82, CH-4056 Basel, Switzerland}

\author{Kenji Watanabe}
\affiliation{Research Center for Functional Materials, National Institute for Materials Science, 1-1 Namiki, Tsukuba, 305-0044, Japan}

\author{Takashi Taniguchi}
\affiliation{International Center for Materials Nanoarchitectonics, National Institute for Material Science, 1-1 Namiki, Tsukuba, 305-0044, Japan}

\author{Dominik M. Zumbühl}
\affiliation{Department of Physics, University of Basel, Klingelbergstrasse 82, CH-4056 Basel, Switzerland}
\affiliation{Swiss Nanoscience Institute, University of Basel, Klingelbergstrasse 82, CH-4056 Basel, Switzerland}

\author{Christian Sch\"onenberger}
\affiliation{Department of Physics, University of Basel, Klingelbergstrasse 82, CH-4056 Basel, Switzerland}
\affiliation{Swiss Nanoscience Institute, University of Basel, Klingelbergstrasse 82, CH-4056 Basel, Switzerland}



\date{\today}

\begin{abstract}
Two dimensional materials are usually envisioned as flat, truly 2D layers. However out-of-plane corrugations are inevitably present in these materials. In this manuscript, we show that graphene flakes encapsulated between insulating crystals (hBN, WSe2), although having large mobilities, surprisingly contain out-of-plane corrugations. The height fluctuations of these corrugations are revealed using weak localization measurements in the presence of a static in-plane magnetic field. Due to the random out-of-plane corrugations, the in-plane magnetic field results in a random out-of-plane component to the local graphene plane, which leads to a substantial decrease of the phase coherence time. Atomic force microscope measurements also confirm a long range height modulation present in these crystals. Our results suggest that phase coherent transport experiments relying on purely in-plane magnetic fields in van der Waals heterostructures have to be taken with serious care.
\end{abstract}

\pacs{}

\maketitle 

\section{Introduction}
\label{sec:intro}
From thermodynamical considerations it was long thought that two-dimensional (2D) crystals cannot exist at non-zero temperatures \cite{1934_Peierls, 1935_Peierls, 1937_Landau, 1968_Mermin}. Therefore, it came as a big surprise when graphene (Gr)\cite{2004_Novoselov}, a truly single layer of graphite, was first discovered. In the following years several other crystals followed \cite{2005_Novoselov_a} among which the transition metal dichalchogenids (TMDC) are maybe the most famous. The existence of these crystals were attributed to either the presence of the substrate or the appearance of out-of-plane corrugations in the layer\cite{2007_Meyer, 2007_Fasolino}. Both arguments lift effectively the two-dimensionality of the crystals.

Out-of-plane corrugations and ripples have been observed in graphene both by AFM and TEM studies \cite{2007_Meyer, 2007_Ishigami, 2009_Lui}. These can originate from thermodynamic reasons, from straining during exfoliation, or from the underlying substrate corrugations. It was found that graphene placed on  SiO$_2$ conformally deforms, resulting in out-of-plane corrugations \cite{2009_Geringer, 2010_Cullen}. This has a strong impact on the transport properties. On the one hand, it can reduce the mobility by introducing strain-induced scalar and vector potentials which lead to long-range disorder and hence to additional scattering \cite{2008_Katsnelson, 2014_Couto, 2019_Kim, 2019_Wang}. On the other hand, a more direct and striking consequence can be observed if the system is placed into an in-plane magnetic field. Random out-of-plane magnetic field components originate from the corrugated graphene sheet, as shown in Fig.~\ref{fig:device}~(a). The corrugations are described by their root-mean-square (rms) height ($Z$) and correlation length ($R$). The resulting random out-of-plane magnetic fields can be seen as random vector potentials that lead to substantial dephasing in weak localization measurements as demonstrated by Lundeberg and Folk\cite{2010_Lundeberg}.

With the introduction of hexagonal boron nitride (hBN) as a substrate for graphene devices \cite{2010_Dean, 2013_Wang_a}, not only a more silent dielectric, but also an atomically smooth substrate was found. Recently, TMDCs have emerged as an alternative substrate for exceptionally clean graphene devices \cite{2014_Kretinin, 2014_Lu, 2017_Banszerus_a, 2018_Zihlmann, 2019_Banszerus}. However, the remaining mobility limiting disorder has not yet been identified in these vdW-heterostructures. Strong evidence for strain fluctuations as the remaining disorder has been found in single layer graphene\cite{2014_Couto, 2019_Kim} as well as in bilayer graphene devices \cite{2014_Engels}. Furthermore, in our recent study we could in-situ tune and increase the mobility of hBN-encapsulated graphene devices by applying a global uniaxial strain to our heterostructures \cite{2019_Wang}. In doing so, random strain fluctuations are reduced and hence the mobility is increased. In principle, strain fluctuations can be either of in-plane or of out-of-plane nature. However, the fact that they are tunable with global uniaxial strain\cite{2019_Wang} and the fact that they are reduced by AFM ironing\cite{2019_Kim} hints towards out-of-plane corrugations as the dominant source of nanometre strain fluctuations. 

In this work we present phase coherent magnetotransport studies on several devices encapsulated in hBN or between WSe$_2$ and hBN. The measured phase coherence time gives valuable insight into out-of-plane corrugations. By applying an in-plane magnetic field the phase coherence time drops even for devices appearing at first sight to be flat and bubble free. The measurements are well accounted by the model introduced in Ref.~\onlinecite{2001_Mathur} and \onlinecite{2010_Lundeberg} and are also in accordance with detailed AFM studies. Our measurements are an unambiguous proof of the presence of out-of-plane corrugations in vdW heterostructures. These corrugations could be the origin of the random strain fluctuations limiting the charge carrier mobility and could limit phase coherent transport experiment in the presence of an in-plane magnetic field.

\section{Results}
\label{sec:results}
In the following, experimental data of three different vdW-heterostructures is presented. An overview of the three samples is given in table~\ref{tab:devices} in the appendix.

\begin{figure}[htbp]
	\centering
	\includegraphics[width=8.5cm]{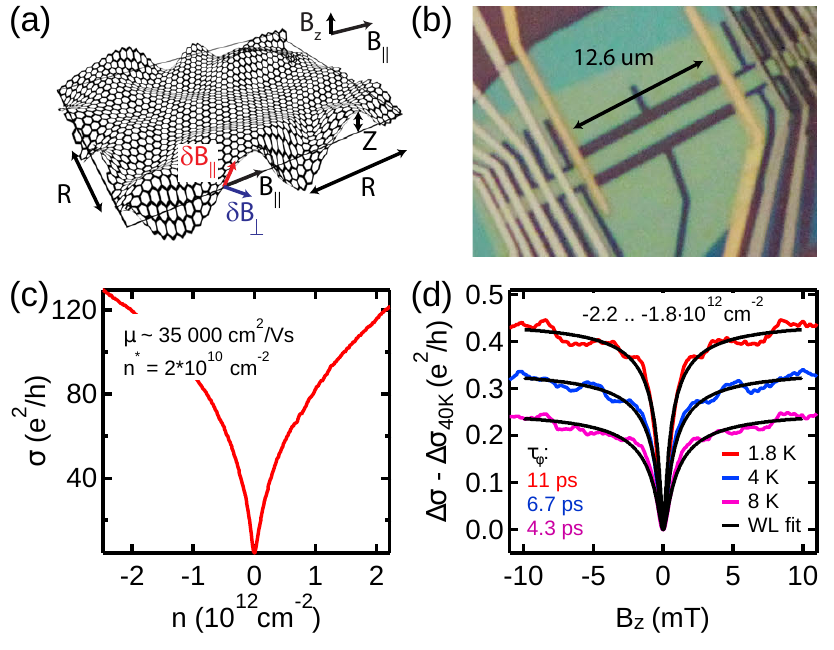}
	\caption{\label{fig:device} \textbf{(a)} Out-of-plane corrugations in a graphene lattice (exaggerated) with lateral correlation length $R$ and rms height $Z$. A uniform in-plane magnetic field $B_\parallel$ (black) will lead to a random surface normal ($\delta B_\perp$, blue) and a parallel ($\delta B_\parallel$, red) component. An homogeneous out-of-plane magnetic field ($B_z$) is used to study phase coherent transport. \textbf{(b)} Optical image of device 1 (hBN/Gr/hBN). \textbf{(c)} Two-terminal conductivity of device 1 as a function of charge carrier density. \textbf{(d)} Ensemble averaged weak localization correction at different temperatures at a hole doping of \SI{-2.2e12}{\square\per\centi\metre}$<n<$\SI{-1.8e12}{\square\per\centi\metre} and zero in-plane magnetic field.}
\end{figure}

An optical image of device~1 after fabrication is shown in Fig.~\ref{fig:device}~(b). It is a two terminal hBN/Gr/hBN heterostructure with a large aspect ratio (length/width), which makes it ideal for magnetoconductance measurements. The conductivity as a function of charge carrier density is shown in Fig.~\ref{fig:device}~(c), from which we extract a mobility of \SI{35000}{\square\centi\metre\per\volt\per\second} and a residual doping of \SI{2e10}{\square\per\centi\metre}. At low temperature, phase coherent transport leads to weak localization as shown in Fig.~\ref{fig:device}~(d). Here, we plot an ensemble averaged quantum correction to the magneto conductivity at different temperatures, which we obtain by subtracting the classical magneto conductivtiy measured at \SI{40}{K} from the low temperature measurements. The quantum correction to the magnetoconductivity can be fitted by the standard weak localization (WL) formula for graphene\cite{2006_McCann}:
\begin{equation}
	\label{eq:WL}
	\begin{split}
	\Delta\sigma (B) = \frac{e^2}{\pi h}\left[ F \left( \frac{\tau_B^{-1}}{\tau_\phi^{-1}} \right) - F \left( \frac{\tau_B^{-1}}{\tau_\phi^{-1} + 2\tau_{iv}^{-1}} \right) \right. \\
	\left. - 2F \left( \frac{\tau_B^{-1}}{\tau_\phi^{-1} + \tau_{iv}^{-1} + \tau_{*}^{-1}} \right) \vphantom{\int_1^2} \right],
	\end{split},
\end{equation}
where $F\left(x\right) = \ln\left(x\right) + \Psi\left(1/2 + 1/x\right)$, with $\Psi\left(x\right)$ being the digamma function, $\tau_B^{-1} = 4DeB/\hbar$, where $D$ is the diffusion constant, $\tau_\phi$ the phase coherence time, $\tau_{iv}$ the intervalley scattering time and $\tau_*$ the intravalley scattering time. We fit the curves corresponding to the three different temperatures within the same fitting procedure (global fit), where only $\tau_{\phi}$ is allowed to change with temperature to extract all relevant scattering time scales\cite{2008_Tikhonenko}. We generally find phase coherence times on the order of a few pico seconds at a temperature of a few Kelvins and an intervalley scattering time $\tau_{iv}$ of \SI{6.9e-12}{s} and very small intravalley scattering times $\tau_*\lesssim$~\SI{1e-13}{s}. Whereas the extraction of inter and intravalley scattering times might be challenging in certain cases, the phase coherence time can be reliably extracted in all cases since it is given by the curvature of the magnetoconductivty at zero out-of-plane magnetic field.


In order to probe the out-of-plane corrugations we study phase coherent transport as a function of small out-of-plane magnetic fields in the presence of large, static in-plane magnetic fields. It is obvious that a homogeneous in-plane magnetic field leads to random out-of-plane components $\delta B_\perp$ if the graphene sheet has out-of-plane corrugations, see Fig.~\ref{fig:device}~(a). These random out-of-plane magnetic field components can be described as random vector potentials that affects phase coherent transport.

First experiments have been realized on Si inversion layers\cite{1987_Mensz, 1993_Anderson} and two-dimensional electron gases in GaAs-heterojunctions\cite{1987_Mensz}. A direct correlation between the topographic morphology and the dephasing rate has been found. Motivated by these findings, Mathur and Baranger have calculated the additional dephasing for a two-dimensional electron gas originating from Gaussian correlated corrugations in the presence of an in-plane magnetic field\cite{2001_Mathur}:
\begin{equation}
	\label{eq:dephasing}
	\tau_\phi^{-1} \rightarrow 	\tau_\phi^{-1} + \sqrt{\pi} \frac{e^2}{\hbar^2}vZ^2RB_\parallel^2.
\end{equation}
Here, $Z$ is the root-mean-square of the corrugation height, $R$ the lateral correlation length of the corrugations, $B_\parallel$ is the in-plane magnetic field, $v$ the Fermi velocity and constants $e$ and $\hbar$ are the electronic charge and the reduced Planck constant respectively.

\begin{figure}[htbp]
	\centering
	\includegraphics[width=8.5cm]{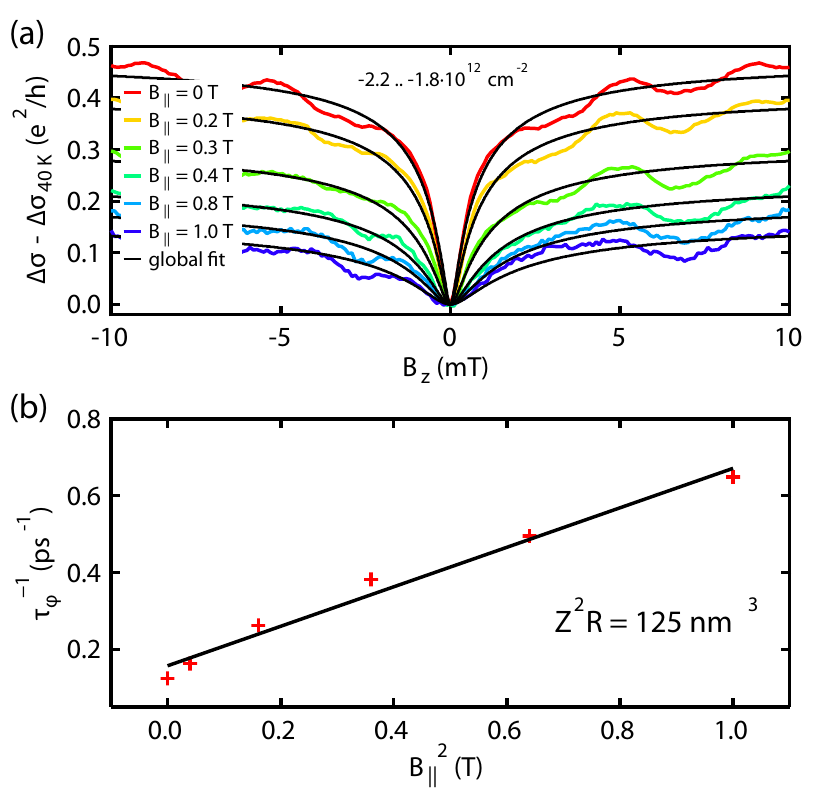}
	\caption{\label{fig:dev1} \textbf{(a)} WL of device 1 for different values of in-plane field $B_\parallel$ at a temperature of \SI{1.8}{K} . The fitted dephasing rate $\tau_\phi^{-1}$ as a function of $B_\parallel^2$ is shown in \textbf{(b)}.}
\end{figure}

The quantum correction to the magnetoconductivity for different in-plane magnetic fields is shown in Fig.\ref{fig:dev1}~(a). Here we show a representative data set for hole doping. Single WL curves represent an ensemble averaged measurement over a density range of \SIrange{-2.2e12}{-1.8e12}{\square\per\centi\metre}. Similar effects have been observed for electron doping. Different colours represent WL curves measured at different in-plane magnetic fields. As the in-plane field increases, the dip around zero magnetic field gets less pronounced and the overall magnetoconductance reduces. This changes can be understood by a reduced phase coherence time $\tau_\phi$.

In order to perform a quantitative analysis, we apply a global fit (fitting all the curves with different $B_\parallel$ at the same time) where only  $\tau_{\phi}$ is allowed to vary since neither  $\tau_{iv}$ nor  $\tau_{*}$ are expected to be affected by $B_\parallel$. The extracted dephasing rate $\tau_\phi^{-1}$ is shown in Fig.~\ref{fig:dev1}~(b) as a function of $B_\parallel^2$. A clear linear behaviour is observed that allows us to extract the corrugation volume $Z^2R=$\SI{125}{nm^3} using Eq.~\ref{eq:dephasing}. The relatively large corrugation volume, which is much larger than the previously reported value of \SI{1.7}{nm^3} for graphene on SiO$_2$ \cite{2010_Lundeberg}, can be explained by the presence of bubbles in the device. Bubbles with contaminations are known to spontaneously form at interfaces of vdW heterostructures \cite{2012_Haigh, 2014_Kretinin} and were confirmed by optical and AFM images of device 1. Therefore, it is not surprising that an additional dephasing is observed when device 1 is placed in an in-plane magnetic field. 

\begin{figure}[htbp]
	\centering
	\includegraphics[width=8.5cm]{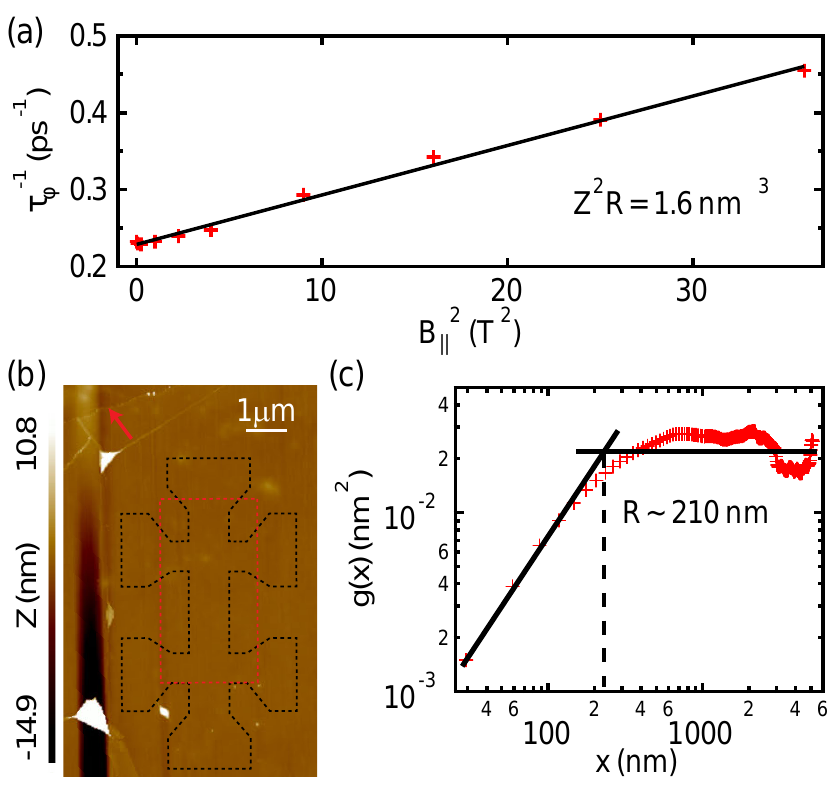}
	\caption{\label{fig:dev2} \textbf{(a)} Fitted dephasing rate of device 2 as a function of $B_\parallel^2$. A ripple volume of \SI{1.6}{nm^3} is extracted. \textbf{(b)} AFM image of the graphene flake on top of the bottom hBN crystal. The graphene flake outline is marked by the red arrow. The location of the Hall bar is indicated by the black dashed line and the area used for the AFM analysis is highlighted by a red dashed rectangle. \textbf{(c)} Height-height correlation of device 3 extracted from the AFM measurement in (b). A correlation length of \SI{~210}{nm} and an average height fluctuation of \SI{96\pm 1.2}{pm} is found.}
\end{figure}

However, we also found an additional dephasing in much cleaner and essentially bubble free hBN/Gr/hBN heterostructures. We have measured the phase coherence time as a function of in-plane magnetic field $B_\parallel$. Here we directly show in Fig.~\ref{fig:dev2}~(a) the extracted dephasing rate as a function of $B_\parallel^2$ of device 2 at a doping of \SIrange{0.3e12}{1.2e12}{\square\per\centi\metre}. The dephasing rate was extracted from the curvature of the magnetoconductivity at zero out-of-plane magnetic field, see Fig.~\ref{fig:app:device3electorn} in appendix~\ref{app:sec:magnetoconductance_mirko} for more details. This device, which is free from bubbles (confirmed by AFM measurements, see Fig.~\ref{fig:dev2}~(b)), shows a corrugation volume $Z^2R=$~\SI{1.6}{nm^3}. This is two orders of magnitude smaller than the extracted volume for device 1 that contains bubbles and surprisingly close to the corrugation volume of graphene on SiO$_2$.

The additional dephasing in an in-plane magnetic field only gives access to the total corrugation volume $Z^2R$ and not to the individual contributions of height $Z$ and radius $R$. We used high resolution AFM images to extract the standard deviation of a Gaussian height distribution that corresponds to the corrugation height $Z$ and the height-height correlation length that corresponds to the corrugation radius $R$. An AFM image of device 2, with the outline of the Hall bar, before placing the top hBN, is shown in Fig.~\ref{fig:dev2}~(b)(see appendix for further fabrication details). From the height distribution, we extracted $Z$=\SI{96\pm 1.2}{pm}, see Fig.~\ref{fig:height_distribution} in appendix~\ref{app:AFM}. In addition, the same dataset is used to extract the height-height correlation length, which is the characteristic length scale for the corrugations. As shown in Fig.~\ref{fig:dev2}~(c), the correlation length $R$ corresponds to the crossover between the small and large length scale behaviour of the correlation function as evident in a log-log plot. We find $R\sim$~\SI{210}{nm} for this device. Analysis of further graphene/hBN half-stacks prepared in the same manner, revealed very similar values for the height distribution and lateral correlation length. Thus, we find a corrugation volume by analysing AFM images $Z^2R_{AFM}\sim$~\SI{1.9}{nm^3}. This independent rough estimation of the corrugation volume matches the corrugation volume extracted from transport measurements well. 

Even though graphene is sandwiched between two layers of atomically flat hBN crystals, out-of-plane corrugations are present. The corrugation volume in bubble free hBN encapsulated graphene (\SI{1.6}{nm^3}) is similar to the corrugation volume of SiO$_2$ supported graphene (\SI{1.7}{nm^3}). However, in the case of hBN encapsulated graphene, the corrugations have a smaller height but larger lateral extension compared to the relatively short length scales in graphene on SiO$_2$, which is on the order of a few nanometers \cite{2010_Lundeberg, 2007_Ishigami}.

\begin{figure}[htbp]
	\centering
	\includegraphics[width=8.5cm]{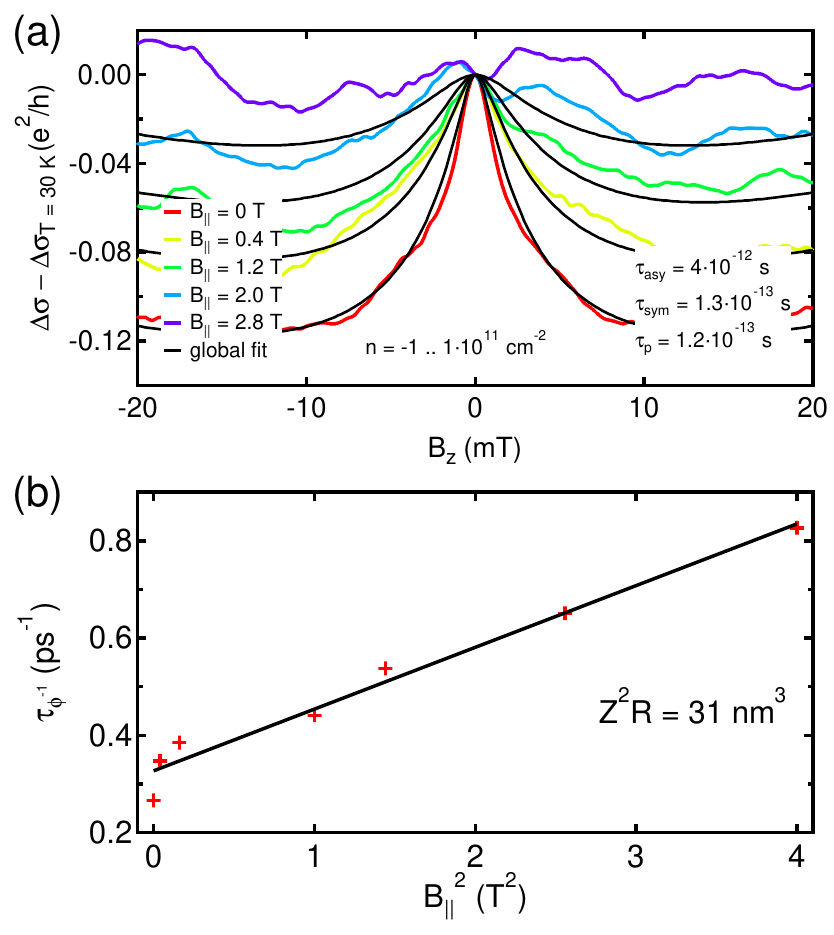}
	\caption{\label{fig:dev3} \textbf{(a)} WAL of device 3 for different values of in-plane magnetic field $B_\parallel$ at at temperature of \SI{1.8}{K}. The fitted dephasing rate $\tau_\phi^{-1}$ as a function of $B_\parallel^2$ is shown in \textbf{(b)}.}
\end{figure}

Out-of-plane corrugations are not limited to hBN encapsulated graphene, but are a generic phenomena in vdW-hetersotructures. Here we present phase coherent transport in hBN/Gr/WSe$_2$ heterostructures, where additional dephasing is observed when an in-plane magnetic field is applied. Fig.~\ref{fig:dev3}~(a) shows the quantum correction of the magneto conductivity that exhibits weak anti-localization (WAL) due to graphene's proximity to the TMDC WSe$_2$ \cite{2018_Zihlmann}. In the case of graphene, the quantum correction to the magneto conductivity $\Delta\sigma$ in the presence of strong SOC is given by \cite{2012_McCann}:
\begin{equation}
	\label{eq:WAL}
	\begin{split}
	\Delta\sigma (B) = -\frac{e^2}{2\pi h}\left[ F \left( \frac{\tau_B^{-1}}{\tau_\phi^{-1}} \right) - F \left( \frac{\tau_B^{-1}}{\tau_\phi^{-1} + 2\tau_{asy}^{-1}} \right) \right. \\
	\left. - 2F \left( \frac{\tau_B^{-1}}{\tau_\phi^{-1} + \tau_{asy}^{-1} + \tau_{sym}^{-1}} \right) \vphantom{\int_1^2} \right],
	\end{split}
\end{equation}
where $F\left(x\right) = \ln\left(x\right) + \Psi\left(1/2 + 1/x\right)$, with $\Psi\left(x\right)$ being the digamma function, $\tau_B^{-1} = 4DeB/\hbar$, where $D$ is the diffusion constant, $\tau_\phi$ the phase coherence time, $\tau_{asy}$ ($\tau_{sym}$) the spin-orbit scattering time that takes only spin-orbit terms into account that are asymmetric (symmetric) in $z\rightarrow -z$ direction. Assuming that $\tau_{asy}$ and $\tau_{sym}$ are independent of $B_\parallel$, we perform a global fit where only $\tau_\phi$ is allowed to vary with $B_\parallel$. We have also varied the spin orbit times, which did essentially not change within the limits of the extraction method, see also Ref.~\onlinecite{2018_Zihlmann} for further information. The extracted dephasing rate scales linearly as a function of $B_\parallel^2$ as shown in Fig.~\ref{fig:dev3}~(b). Using equation~\ref{eq:dephasing}, a corrugation volume $Z^2R=$\SI{31}{nm^3} is extracted. This is roughly an order of magnitude smaller (larger) than in device 1 (2).

\section{Discussion}
\label{sec:discussion}
Despite the fact that hBN crystals are atomically flat and of high quality, graphene encapsulated between two such crystals exhibits out-of-plane corrugations. The corrugation volume extracted from phase coherent transport measurements varies among different vdW heterostructures depending on their interface and crystal quality. The large corrugation volume observed in device 1 is obviously originating from bubbles at the Gr/hBN interface and can be avoided by utilizing heterostructures with a better interface quality or by designing the active area of the device in a bubble free part of the heterostructure. However, in essentially bubble free heterostructures, out-of-plane corrugations are still present. Whereas for device 3 the vdW-interface might still be the limiting factor in terms of origin of out-of-plane corrugations, this is certainly not the case in device 2 where AFM images show a smooth, atomically flat surface without any bubbles or contaminations. Here, a different explanation for the presence of out-of-plane corrugations has to be invoked. One possibility is that the crystal quality of hBN might influence the remaining out-of-plane corrugations since defects in the hBN crystal (also in layers far away from the interface to the graphene) might lead to long-range height fluctuations\cite{2017_Rooney}. Moreover, residues trapped at the bottom hBN - SiO$_2$ interface might lead to long range height fluctuations at the top of the bottom hBN.

Our results are in agreement with previous measurements on graphene on SiO$_2$ substrate \citep{2010_Lundeberg}. It is clear that devices with bubbles exhibit a larger corrugation volume than devices on a clean SiO$_2$ surface. However, it is surprising that not even the best hBN/Gr/hBN devices show a smaller corrugation volume than devices on SiO$_2$ substrates. It is important to note, that even though the corrugation volume is similar, the lateral correlation length is much longer and the height variation is considerably smaller for vdW heterostructures ($Z\sim$~\SI{0.1}{nm}, $R\sim$~\SI{200}{nm}) compared to a SiO$_2$ substrate ($Z\sim$~\SI{0.4}{nm}, $R\sim$~\SI{5}{nm}\cite{2009_Geringer, 2010_Cullen}). Therefore, the deformations of the graphene lattice and hence the random strain fluctuations are greatly reduced in graphene in vdW-heterostructures compared to graphene on SiO$_2$ substrates. This might be one of the reasons why the graphene quality in fully encapsulated graphene can be exceptionally good.

Finally, we would like to raise the point that the presence of out-of-plane corrugations (even in the cleanest devices) might impose severe limitations on phase coherent experiments relying on large in-plane magnetic fields. This for example prevents one from studying the transition from WAL to WL in graphene with spin-orbit coupling triggered by an in-plane magnetic field \cite{2012_McCann, 2018_Zihlmann}. In addition, the magnitude of the supercurrent in a graphene based Josephson junction in an in-plane magnetic field could be reduced due to a reduced phase coherence time.

\section{Conclusion}
\label{sec:conclusion}
In conclusion, phase coherent transport has shown that out-of-plane corrugations are present in vdW-heterostructures. The corrugation volume strongly depends on the interface quality between graphene and other 2D-material (e.g. hBN and WSe$_2$) but is non-zero even for the best interface and device quality ($\mu\sim$~\SI{100000}{\square\centi\metre\per\volt\per\second}). The presence of out-of-plane corrugations implies distortions of the graphene lattice and hence also random strain fluctuations. While the corrugation volume for the cleanest hBN/Gr/hBN device is similar to graphene on SiO$_2$, its effect on transport is greatly reduced because of the long range nature of the corrugations in hBN/Gr/hBN (smaller strain fluctuations). Nonetheless, phase coherent experiments relying on large in-plane magnetic fields could suffer from the out-of-plane corrugations due to a reduced phase coherence time.


%
%

%

\begin{acknowledgments}
This work has received funding from the European Union’s Horizon 2020 research and innovation programme under grant agreement 696656 (Graphene Flagship), 787414 (ERC-Adv TopSupra), and 824109 (European Microkelvin Platform EMP); the Swiss National Science Foundation (including 179024); the Swiss Nanoscience Institute; the Swiss NCCR QSIT; and Topograph FlagERA network OTKA FK-123894. This research was supported by the National Research, Development and Innovation Fund of Hungary within the Quantum Technology National Excellence Program (Project Nr.  2017-1.2.1-NKP-2017-00001). P.M. acknowledges support from the Marie Curie and Bolyai fellowships. K.W. and T.T. acknowledge support from the Elemental Strategy Initiative conducted by the MEXT, Japan ,Grant Number JPMXP0112101001,  JSPS KAKENHI Grant Numbers JP20H00354 and the CREST(JPMJCR15F3), JST.

\end{acknowledgments}

\textbf{Author contributions}

Devices were fabricated by S.Z., M.K.R, D.I. and M.K. Measurements were performed by S.Z. and M.K.R with the help of P.M.. S.Z. and M.K.R. analysed the data with help from P.M. and inputs from C.S. and D.M.Z.. S.Z., P.M., L.W., D.I. and C.S. were involved in the interpretation of the results. S.Z. and P.M. co-wrote the manuscript with inputs from all authors. D.M.Z, C.S. and P.M. guided the work. K.W. and T.T. provided the hBN crystals used in the devices.

\newpage
\appendix

\section{Overview of the measured devices}
\label{app.sec:devices}
Table~\ref{tab:devices} shows an overview of the three devices.

\setlength{\tabcolsep}{8pt}
\begin{table*}[hbtp]
	\centering
	\renewcommand{\arraystretch}{1.3}
	\caption{\label{tab:devices} Overview of the three devices. $D, l_{mfp}, \tau_\phi, l_\phi, B_{tr}$ are given for the density range that was used for the evaluation of the out-of-plane corrugations in the density interval ($n_{eval}$).}
    \begin{tabular}{llllllllll}
    \toprule[1.5pt]
        ~   	& $\mu$	& $n_{eval}$ &$D$& $l_{mfp}$  & $\tau_\phi$ & $l_\phi$ & $B_{tr}$ & $Z^2R_{tr}$ & $Z^2R_{AFM}$ \\  
         	& cm$^2$/Vs & 10$^{12}$cm$^{-2}$ & m$^2$/s & \si{nm}& \si{ps}& \si{\micro\meter} & \si{mT} & \si{nm^3}& \si{nm^3}\\ \hline
        device 1  & \num{35000} & -2.2..-1.8 & 0.32& 640 & 8 & 1.6 &\num{~10}& \num{125}  & na  \\
        hBN/Gr/hBN & & & & & & & &  \\ 
        device 2   & \num{120000} & 0.3..1.2 & 0.23& 460  & 4.3 & \num{0.98} & \num{~20}& \num{1.6}&  \num{1.9}  \\ 
        hBN/Gr/hBN & & & & & & & &  \\ 
        device 3  & \num{130000} & -0.1..0.1 & 0.075& 150  & 3.8 & 0.53 & \num{~180}& \num{31} & na  \\
        WSe$_2$/Gr/hBN & & & & & & & &  \\ 
    \bottomrule[1.5pt]
    \end{tabular}
\end{table*}

\subsection*{Fabrication of device 1 and device 3}
The vdW-heterostructures of device 1 and device 3 were assembled using a dry pick-up method \cite{2014_Zomer} and Cr/Au 1D-edge contacts were used\cite{2013_Wang_a}. After shaping the vdW-heterostructure into a Hall-bar geometry by a reactive ion etching plasma employing SF$_6$ as the main reactive gas, Ti/Au top gates with an MgO dielectric layer were fabricated on device 3. A heavily-doped silicon substrate with \SI{300}{\nano\metre} SiO$_2$ was used as a global back gate for both devices.

\subsection*{Fabrication of device 2}
Device 2 was not fabricated using the dry pick-up method but relying on a wet process where the graphene is transferred by a PMMA membrane on a hBN flake \cite{2010_Dean}. After PMMA removal in acetone, the sample was annealed at \SI{450}{\celsius} in a hydrogen atmosphere (\SI{1.7}{mbar}). Clean and wrinkle-free areas were identified imaging the the heterostructure by non-contact AFM prior to the deposition of a top hBN layer to protect the graphene from further fabrication steps. These AFM images were also used to extract the data shown in Fig.~\ref{fig:dev2}. The deposition of the top hBN could in principle induce additional corrugation. However, The device shaping as well as the contact deposition was performed in the same way as for device 1 and device 3.

\subsection*{Measurements}
Standard low frequency lock-in techniques were used to measure two- and four-terminal conductances and resistances. Weak (anti-)localization was measured at a temperature of \SI{1.8}{\kelvin} whereas a classical background was measured at sufficiently large temperatures of \SIrange{30}{50}{\kelvin} for device 1 and device 3. Device 2 was measured at \SI{30}{\milli\kelvin} and no classical background was subtracted.
A vector magnet was used to independently control the in-plane ($B_\parallel$) and out-of-plane ($B_z$) magnetic field component.

The quantum correction to the magnetoconductivity was only analysed in the interval $|B_z|\leq B_{tr}$, where $B_{trans} = \Phi_0/l_{mfp}$ is the so-called transport field that describes the limit of diffusive transport. 

The diffusion constant was calculated in the following way:
\begin{equation}
	\label{eq:Einstein_relation}
	D(n) = \frac{\hbar v_F\sqrt{\pi}}{2e^2} \frac{\sigma(n)}{\sqrt{\sqrt{n^2 + n_*^2}}},
\end{equation}
where $\hbar$ is the reduced Planck constant, $v_F = $~\SI{1e6}{m/s} the Fermi velocity of graphene, and $e$ the fundamental unit of charge. The conductivity $\sigma$ was measured and $n$ was calculated from a parallel plate capacitor model. The residual doping, $n^*$, was used as a cut-off to calculate $D$ around the charge neutrality point.

\section{Magneto conductivity of device 2}
\label{app:sec:magnetoconductance_mirko}

\begin{figure}[htbp]
	\centering
	\includegraphics[width=8.5cm]{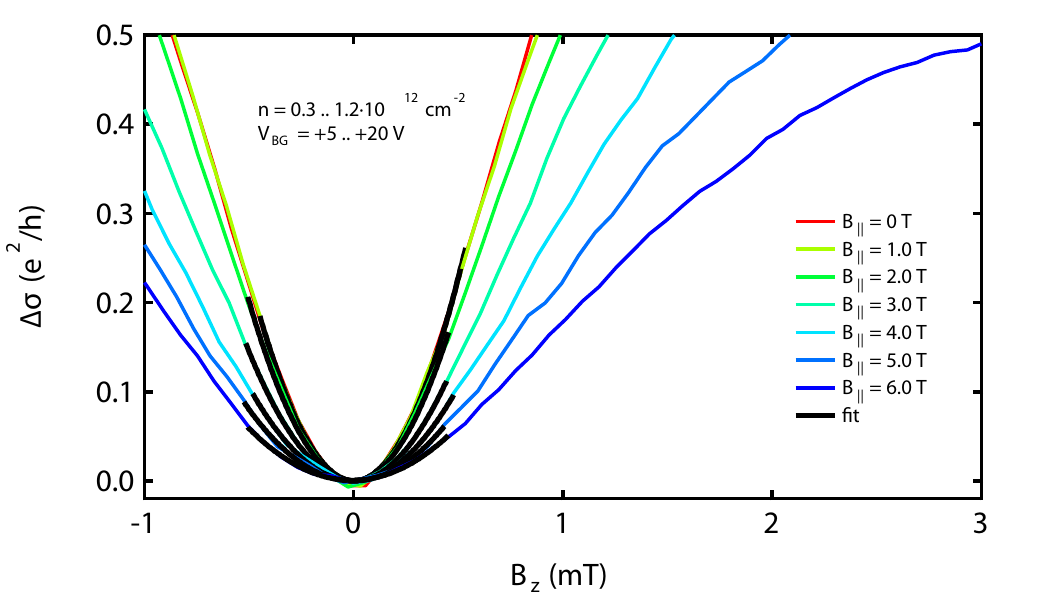}
	\caption{\label{fig:app:device3electorn} Magneto conductivity for various in-plane magnetic fields for electron doping of device 2. The black curves are fits with a parabola to extract the phase coherence time using eq.~\ref{eq:curvature}.}
\end{figure}

Instead of fitting the full WL formula \ref{eq:WL} to the magneto conductivity, the phase coherence time $\tau_\phi$ can also be extracted from the curvature of the magnetoconductance ($\sigma$) at zero out-of-plane magnetic field ($B_\perp$) \cite{2015_Lara-Avila}:
\begin{equation}
	\label{eq:curvature}
	\left.\frac{\partial^2\sigma}{\partial B_\perp^2}\right|_{B_\perp=0} = \frac{16 \pi}{3}\frac{e^2}{h}\left(\frac{D\tau_\phi}{h/e}\right),
\end{equation}
where $D$ is the diffusion constant, $\tau_\phi$ is the phase coherence time and the constants $e$ and $h$ are the electron's charge and Planck's constant respectively. This is especially useful for very high mobility devices where $B_{tr}$ is very small. Fig.~\ref{fig:app:device3electorn} shows the magneto conductivity for electron doping for various in-plane magnetic fields. It is clearly observable that the curvature at zero $B_z$ gets smaller for larger in-plane magnetic field, hence the phase coherence time is smaller as well.

\begin{figure}[htbp]
	\centering
	\includegraphics{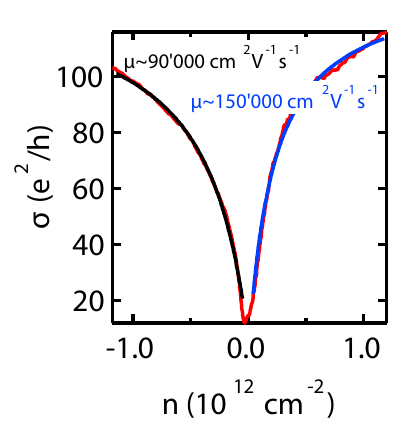}
	\caption{\label{fig:app:device3mobility} Two-terminal conductivity as a function of charge carrier density of device 2. A two-parameter model is fit to extract charge carrier mobility and series resistance (\SI{\sim 190}{\ohm}).}
\end{figure}

\begin{figure}[htbp]
	\centering
	\includegraphics[width=8.5cm]{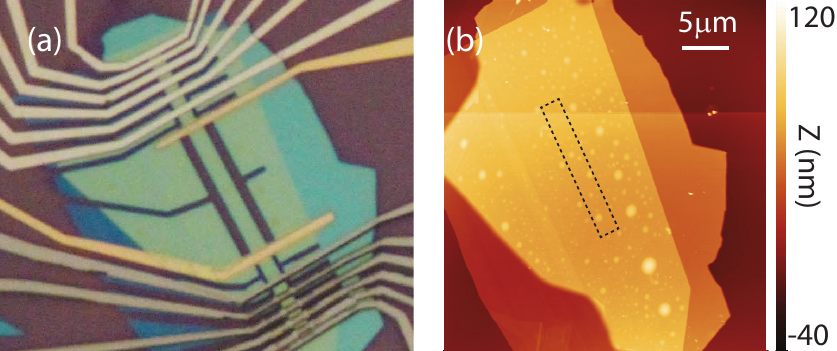}
	\caption{\label{fig:device1_image_AFM}\textbf{(a)} Optical image and \textbf{(b)} AFM image of the full vdW-heterostructure of device 1 with device outline overlaid.}
\end{figure}

\begin{figure}[htbp]
	\centering
	\includegraphics[width=8.5cm]{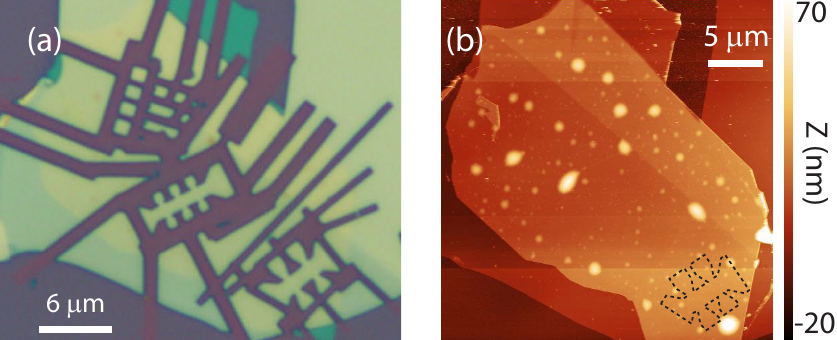}
	\caption{\label{fig:device3_image_AFM}\textbf{(a)} Optical image and \textbf{(b)} AFM image of the full vdW-heterostructure of device 3 that shows that the device area is essentially bubble free. Data presented in the main text has been taken from the Hall bar outlined in \textbf{(b)}.}
\end{figure}

\newpage
\section{Additional information on the AFM analysis}
\label{app:AFM}
The AFM analysis presented in the main text and below has been performed on device 2 before placing the top hBN. The area marked by the red dashed rectangle shown in Fig.~\ref{fig:dev2}~(b), which corresponds to the area where the Hall bar has been defined, was used to extract $Z$ and $R$. Therefore, the AFM analysis and the transport measurements probe the same area. A tilted plane has been subtracted from the height data prior to the detailed analysis described in the following.

\begin{figure}[htbp]
	\centering
	\includegraphics[width=8.5cm]{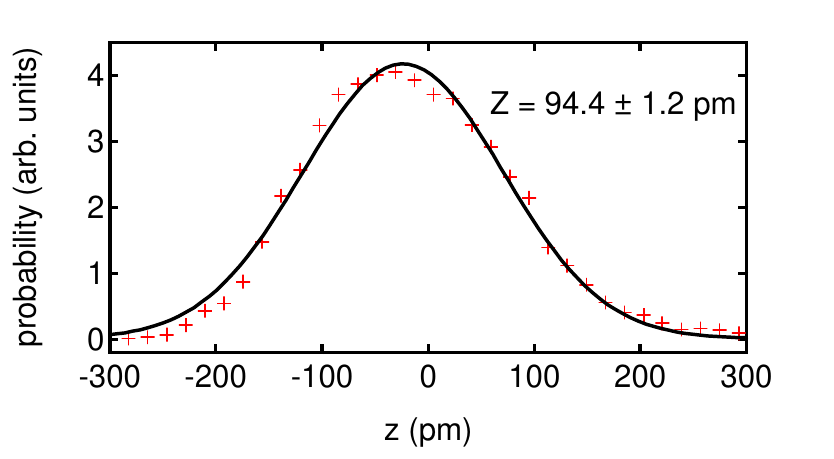}
	\caption{\label{fig:height_distribution} The height distribution (red crosses) evaluated in the area marked by the red dashed rectangle in Fig.~\ref{fig:dev2}~(b). A Gaussian fit (black solid line) yields Z = \SI{94.4\pm1.2}{pm}.}
\end{figure}

The rms height ($Z$) of the corrugations can directly be extracted from AFM measurements by either calculating it from the raw data (point by point) or by fitting the height distribution with a Gaussian model. The point by point calculation of $Z$ is given as:
\begin{equation}
	\label{eq:rms}
	Z = \frac{1}{N}\sum_{n=1}^N \left(z_n - \bar{z}\right)^2,
\end{equation}
where $z_n$ is the height value of point $n$ and $\bar{z}$ is the average height value of all $N$ points of the AFM image. Additionally, the height distribution as shown in Fig.~\ref{fig:height_distribution}, can be used to extract $Z$. We find $Z$~=~\SI{94.4\pm 1.2}{pm} from the width of the height distribution, which is in good agreement with literature values for graphene on hBN\cite{2010_Dean}.

The height-height correlation function in one dimension $g(x) = \langle\left(z\left(x_0 + x\right) - z\left(x_0\right)\right)^2\rangle$\cite{2007_Ishigami} is defined as follows for a discrete dataset:
\begin{equation}
	\label{eq:HHCF_discrete}
	g\left(x\right) = \frac{1}{N(M-m)}\sum_{l=1}^{N}\sum_{n=1}^{M-m}\left(z_{n+m, l} - z_{n,l}\right)^2,
\end{equation}
where $m = x/\Delta x$ and $\Delta x$ is the spacing between two points. Eq.~\ref{eq:HHCF_discrete} represents the 1-dimensional height-height correlation function averaged over $N$ lines of the second lateral dimension as commonly used in the analysis of AFM images, where $g\left(x\right)$ is calculated for the fast scanning directions and averaged over the slow scanning direction. Eq.~\ref{eq:HHCF_discrete} has been used to calculate $g\left(x\right)$, which is shown in Fig.~\ref{fig:dev2}~(c) in the main text. The correlation length $R$ is identified as the crossover from the polynomial short range behaviour to the constant long range behaviour.

For a Gaussian correlated surface, as assumed by Mathur and Barangar\cite{2001_Mathur} for their calculation, the height-height correlation function takes the form:
\begin{equation}
	\label{eq:HHCF_gauss}
	g(x) = 2Z^2\cdot\left(1 -  e^{-x^2 / R^2}\right),
\end{equation}
where $Z$ is the root mean square deviation from the mean height and $R$ is the correlation length. This is a direct result of the assumption of a Gaussian correlated surface that is defined by a height distribution $z\left(x\right)$ with zero mean and a variance given by
\begin{equation}
	\label{eq:corrugated_surface}
	\langle z\left(x\right) z\left(x\prime \right)\rangle = Z^2 \cdot e^{-\left(x - x\prime\right)^2/R^2},
\end{equation}
where $x$ and $x\prime$ are positions along the x-direction, $Z$ is the rms height fluctuation and $R$ is the correlation length.

The crossover between the short range behaviour and the long range behaviour takes place at $R\sim$~\SI{230}{nm}, as indicated by the crossing of the black dashed lines in Fig.~\ref{fig:dev2}(c) in the main text. In addition, the value for the long range behaviour ($2Z^2$ as given by eq.~\ref{eq:HHCF_gauss} for large, uncorrelated distances) agrees well with $Z$ extracted from the height distribution. Fitting the data with eq.~\ref{eq:HHCF_gauss} results in  $Z=$~\SI{106\pm 1}{pm} and $R=$~\SI{187\pm 20}{nm}, which gives a reasonable agreement with the general calculation. 


We would like to note that the corrugation volume extracted from transport measurements can directly be compared to the corrugation volume extracted from AFM measurements, see Ref.~\onlinecite{2001_Mathur, 1993_Anderson, 2004_Minkov} for further information. In order that eq.~\ref{eq:dephasing} is valid, it is assumed that the correlation length ($R$) is of short range compared to the mean free path $l_{mfp}$\cite{2001_Mathur}. This condition is fulfilled for device 2 where the correlation length has been determined by AFM measurements. It is a question whether this condition is valid for device 1 and 3. However, the experimental finding ($\tau_\phi^{-1} \propto B_\parallel^2$) suggests also a homogeneous broadening described by eq.~\ref{eq:dephasing}.

%

\bibliography{literature_WL}

\end{document}